\documentclass[aps,prl,twocolumn,amsmath,amssymb]{revtex4}

\usepackage{epsfig}
\usepackage{graphicx}
\usepackage{amsfonts}
\usepackage{amssymb}
\usepackage{amsmath}
\usepackage{amsfonts}
\usepackage{bm}

\renewcommand{\vec}[1]{{\mathbf #1}}

\bibpunct{(}{)}{;}{s}{,}{,}

\def\be{\begin{equation}}
\def\ee{\end{equation}}
\def\bea{\begin{eqnarray}}
\def\eea{\end{eqnarray}}
\begin{document}

\title{Disorder Induced Phase Transitions of Type-II Weyl Semimetal}

\author{Moon Jip Park$^{1,2}$}
\author{Bora Basa$^{3}$}
\author{Matthew J. Gilbert$^{2,3}$}
\affiliation{$^1$ Department of Physics, University of Illinois, Urbana, IL 61801}
\affiliation{$^2$ Micro and Nanotechnology Laboratory, University of Illinois, Urbana, IL 61801}
\affiliation{$^3$ Department of Electrical and Computer Engineering, University of Illinois, Urbana, IL 61801}

\date{\today}

\begin{abstract}
Weyl semimetals are a newly discovered class of materials that host relativistic massless Weyl fermions as their low-energy bulk excitations. Among this new class of materials, there exist two general types of semimetals that are of particular interest: type-I Weyl semimetals, that have broken inversion or time-reversal symmetry symmetry, and type-II Weyl semimetals, that additionally breaks Lorentz invariance. In this work, we use Born approximation to analytically demonstrate that the type-I Weyl semimetals may undergo a quantum phase transition to type-II Weyl semimetals in the presence of the finite charge and magnetic disorder when non-zero tilt exist. The phase transition occurs when the disorder renormalizes the topological mass, thereby reducing the Fermi velocity near the Weyl cone below the tilt of the cone. We also confirm the presence of the disorder induced phase transition in Weyl semimetals using exact diagonalization of a three-dimensional tight-binding model to calculate the resultant phase diagram of the type-I Weyl semimetal.
\end{abstract}
\maketitle

\textit{Introduction}- Weyl semimetals (WSM), which are characterized by the gapless bulk states whose Fermi surfaces are either nodal points or lines, have been an intense area of research\cite{PhysRevB.83.205101,PhysRevB.93.214511,PhysRevLett.107.127205,PhysRevX.5.031013,type2,PhysRevB.86.214514,Murakami,Delplace}. The bulk nodal points of a WSM possess non-degenerate band crossings that are robust under a small perturbations to the Hamiltonian. The low energy excitations of these materials are Weyl fermions that are described by two component spinors. The topological nature of the WSM are revealed through examination of the monopoles of Berry curvature present in the Brillouin zone(BZ) referred to as Weyl nodes. In the WSM, the net monopole charge integrated over the entire BZ is zero as the WSM possesses an equal number of the nodes with the positive and the negative charges\cite{Nielsen1981,Nielsen1983}. In the non-interacting theory, these nodal points can only be eliminated by coming into contact with an oppositely charged Weyl node in momentum space, thereby annihilating one another. WSM have been predicted in a number of different materials and structures each of which is characterized by the symmetries broken. The most studied WSM are type-I WSM (WSM1), characterized by the presence of broken inversion or time-reversal symmetry, and type-II WSM (WSM2), which possess broken Lorentz invariance. Inversion broken WSM1, such as $TaAs$\cite{PhysRevX.5.031013,Xu613}, are characterized by the presence of disconnected Fermi arcs at the surface\cite{PhysRevB.83.205101} that give rise to unconventional transport signatures such as quantum anomalous Hall (QAH) effect\cite{KYang} and the chiral anomaly\cite{nonlocal}. More recently, WSM2, such as $MoTe_2$ and $WTe_2$, have been discovered\cite{MoTe,MoTe2,WTe,WTe2}. WSM2 are characterized by an exotic hyperboloid Fermi surface, where the nodes are tilted in the BZ. Due to the tilted nodes, WSM2 exhibit transport properties that are distinct from the WSM1 including the absence of the chiral anomaly at certain magnetic field angles\cite{fieldselective}, magnetic break down resulting in a collapse of Landau levels\cite{PhysRevLett.116.236401}, and anisotropy of the dynamical conductivity\cite{eightmotion}.

While the gapless states are protected by the underlying topology of the phases, the topology of the band structure may be altered when the finite disorder is present\cite{PhysRevLett.103.196805,PhysRevLett.110.236803,PhysRevLett.114.056801,PhysRevB.85.155138,PhysRevLett.102.136806,PhysRevLett.105.216601}. This is due to the renomalization of the topological mass that determines the band topology. The effect of disorder within topological materials has been studied in a number of different contexts including time-reversal topological insulators\cite{PhysRevLett.103.196805,PhysRevLett.110.236803,PhysRevLett.114.056801,PhysRevB.85.155138,PhysRevLett.102.136806,PhysRevLett.105.216601}, where it is shown that the symmetry preserving disorder may induce a transition between topological insulator and trivial insulator. Furthermore, symmetry preserving disorder may also induce a transition between the weak topological insulator to the strong topological insulator phases. Similarly, the effects of disorder have been examined in WSM1\cite{PhysRevB.93.075108,PhysRevB.93.201302,arix1,PhysRevLett.115.246603} in which WSM1 is shown to undergo a transition between normal insulator(NI) and 3D quantum anomalous hall insulators(QAHI) in the presence of both charge and magnetic disorder.

While the effect of disorder in WSM1 is well-established, the stability of WSM2 in the presence of disorder has not been considered. In this work, we show that WSM1 with small but non-zero tilt undergoes a quantum phase transition to WSM2 when charge and magnetic disorders are present. We illustrate this phase transition by calculating the topological mass renormalization that occurs within the first order Born approximation. We find that the topological mass is renormalized while the tilt of the Weyl cones in remains invariant. As a result, the Fermi velocity near the Weyl cone is also renormalized and leads to the possible phase transition between WSM1 and WSM2. Additionally, we confirm our results using numerical exact diagonalization of a three-dimensional tight binding model. To analyze the effect of the disorder self-energy contribution to the numerically obtained Green's function, we utilize the spectral function, which enables to understand the change of the band structure even in the presence of the disorder\cite{spectral}. Furthermore, we find that the reverse transition from WSM2 to WSM1 is also possible depending on the value and sign of the topological mass. Our work reveals the rich phase diagram of WSM in the presence of disorder and will aid in the experimental characterization of WSM materials.

\textit{Model}-We begin by writing the Hamiltonian a Weyl fermion using the lowest order expansion of the momentum near the Weyl cone as,
\bea
\label{continumm}
H=\sum_{i,j} v_{i,j} k_i \sigma_j+\gamma_{tilt,i} k_i I_2
\eea
where $v_{i,j}$ is a matrix which specifies the Fermi velocity, and $\sigma_i$($I_2$) is the $i$-th pauli matrix and $I_2$ is the $2\times2$ identity matrix. The second term, $\gamma_{tilt,i}$, tilts the cone in $i$-direction and whose presence the breaks Lorentz invariance\cite{type2}. The dispersion of the Hamiltonian in Eq. (\ref{continumm}) is given as,
\bea
E=\gamma_{tilt,i} k_i+ \sqrt{k_i v_{i j} v_{jk} k_k}
\eea
In this work, $v$ is a diagonal matrix corresponding to the velocity in each direction. It should be noted that the choice of the velocity does not effect our conclusions. To consider the disorder effect, we rewrite the Weyl Hamiltonian in the real space in lattice regularized form as,
\begin{equation}
\label{eq:wham}
H_W=H_0+H_{tilt}
\end{equation}
where $H_0$ is the minimal two band model of WSM1, which breaks the time-reversal symmetry\cite{PhysRevLett.107.127205,KYang}, and may be written as
\begin{gather}
H_0=t_x sin(k_x) \sigma_x + t_y sin(k_y) \sigma_y
\\
\nonumber
+[ m_0(2-cos(k_x) -cos(k_y)) +(m_z-t_z cos(k_z))]\sigma_z.
\end{gather}
The second term in Eq. (\ref{eq:wham}) is the tilt term, $H_{tilt}$, that breaks the Lorentz invariance can be generally written up to quadratic order in the lattice regularized form as,
\begin{equation}
H_{tilt}=a_{t,i} sin(k_i) I_2+b_{t,i} cos(k_i) I_2
\end{equation}
Using Eq. (\ref{eq:wham}), WSM2 is characterized by the choice of parameters in which the tilt within any direction dominates the Fermi velocity of $H_0$ near the cone, $|\gamma_{tilt,i}|>|v_{i}|$, where the individual tilts at the cones in the lattice Hamiltonian are given as $\gamma_{tilt}=(a_{t}cos(Q) \pm b_{t} sin(Q))$. In this model, the Weyl cones are located at $Q=(0,0,\pm acos(m_z/t_z))$ in the BZ and the Fermi velocity in each of the cones is given as $v_{x,y}=\frac{\partial E}{\partial k_{x,y}}|_{k=Q}=t_{x,y}$ and $v_{z}=t_z\frac{\partial E}{\partial k_z}|_{k=Q}=t_zsin(Q)$. $m_0$ gaps out the spectrum at $X$ and $Y$ points. Here, we choose the direction of the tilt to be in the $\hat{z}$-direction as the purpose of this work is to observe the phase transition resulting from the renormalized topological mass, $m_z$, in the $z$ direction. As the goal of this work is to examine the effects of disorder on the resultant phases in WSM, a momentum space representation of the Hamiltonian is not useful as the inclusion of disorder forbids the use of a momentum space representation. Therefore, it is necesary to utilize the real space representation of the full Hamiltonian is given as,
\begin{gather}
\label{hdis}
H=\sum_{\delta=\hat{x},\hat{y},\hat{z}}c^\dagger_{(i,j,k)+\vec{\delta}}h_{\vec{\delta}} c_{i,j,k}
 +c^\dagger_{i,j,k}(2m_0I_2+m_z\sigma_z) c_{i,j,k}
 \\
 \nonumber+h.c.
\end{gather}
where $i,j,k$ are the coordinates corresponding to the $\hat{x},\hat{y},\hat{z}$ directions respectively, and $h_{(\hat{x},\hat{y})}=\frac{it_{x,y}}{2}\sigma_{x,y}$, $h_{\hat{z}}=\frac{t_z}{2}\sigma_z+\frac{ia_{t}+b_{t}}{2}I_2$. In the remainder of this work, we place a tilde on the top of each parameter so as to indicate it has been normalized respect to $t_z$.

\begin{figure}
\includegraphics[width=0.5\textwidth]{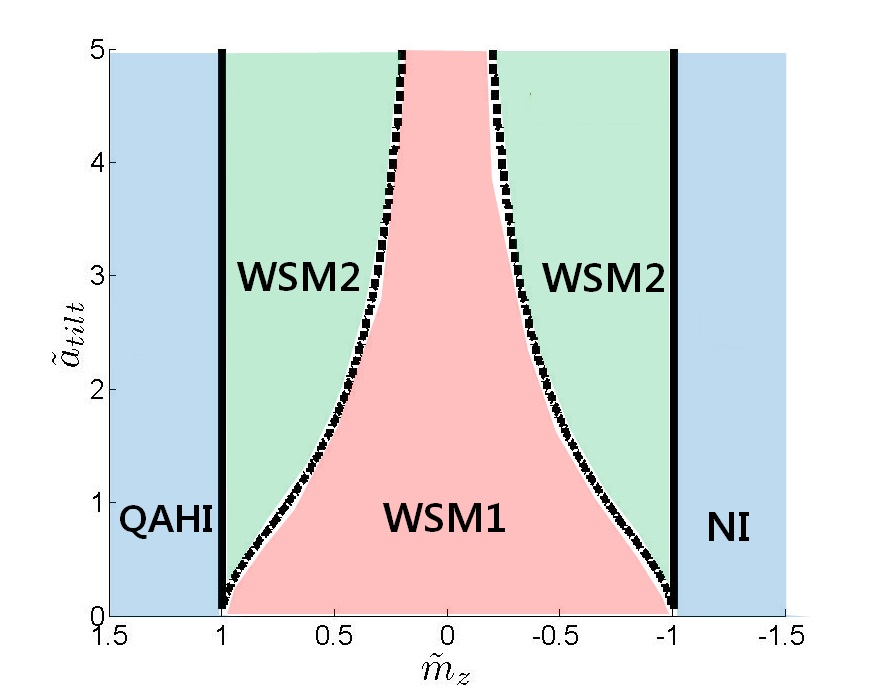}
\caption{\label{fig:cleanphase} Plot of the phase diagram of the WSM in the clean limit. There are the two distinct insulating phase regions corresponding to different values of the topological mass: QAHI when $\tilde{m}_z<-1$ and trivial insulator for $\tilde{m}_z>1$. In this phase diagram, we have ignored $b_{t}$, as the WSM2 transition occurs generally when $\tilde{b}_{t}>1$.
}
\end{figure}

\textit{Phase in the Clean Limit}-In Fig. \ref{fig:cleanphase}, we plot the phase diagram for the WSM as a function of the value of the topological mass and the tilt parameter. The phase diagram of the WSM in the clean limit is characterized by the locations of the Weyl nodes. When a non-zero $m_z$ is present, the Weyl cones are separated in the BZ by the momentum vector $(0,0,2Q)$. As long as the nodes are well-separated in momentum space, the WSM phases are stable in the single particle picture. However, when the nodes with the opposite monopole charge meet one another at the same point in the BZ, they annihilate and produce a gap in the spectrum\cite{chen} resulting in an insulating phase. This transition is mathematically defined as the point in the BZ where $Q$ is ill-defined, namely $\tilde{m}_z>1$ and $\tilde{m}_z<-1$ as is shown in the boundary with the solid lines in Fig. \ref{fig:cleanphase}. There are two distinct insulating phases: the NI phase ($\tilde{m}_z>1$) and QAHI phase ($\tilde{m}_z<-1$). The QAHI phase can be thought as stacks of Chern insulators with a non-zero Hall conductance\cite{PhysRevLett.107.127205}. These two distinct phases of insulator can be heuristically understood by considering a slice of the Hamiltonian at each momentum in $k_z$. In this case, each slice looks like a gapped Dirac fermion except at the gapless nodes at the $(0,0,\pm Q)$. At these points, there exists a non-zero Hall conductance in the interval of $k_z \in (-Q,Q)$, where the outside of the interval the WSM has the zero Hall conductance. As we decrease $m_z$, the Weyl cones shift in momentum space towards the $(0,0,\pm\pi)$ and the region in momentum space with a non-zero Hall conductance extends. Finally, when the Weyl nodes annihilate in the momentum space at $(0,0,\pm\pi)$ which occurs at $\tilde{m_z}=-1$, the gapped bulk is fully specified by the non-zero Hall conductance. In the opposite limit, where the nodes annihilate at $(0,0,0)$ for $\tilde{m_z}=1$, the $k_z$ region with zero Hall conductance extends. In this situation, when the nodes meet in the zero momentum space, the Hamiltonian is characterized by zero Hall conductance resulting in the NI phases.

In addition to the metal-insulator transitions, there exist WSM1-WSM2 transitions as we increase the $m_z$ in the presence of non-zero $\gamma_{tilt}$. This phase transition occurs when the increased $m_z$ lowers the Fermi velocity below the finite value of the tilt term, $\gamma_{tilt}$. This condition can be analytically written for our specific model as $v_f=\tilde{t}_zsin(Q)<\tilde{\gamma}_{tilt}$. In Fig. \ref{fig:cleanphase}, we plot the full phase diagram of the WSM in the clean limit  as a function of the value of $\tilde{\gamma}_{tilt}$ and $\tilde{m}_z$ showing the transitions between WSM1, WSM2, and insulator phases as calculated using the analysis presented.

\textit{The Effect of Disorder}-After establishing the phases of the WSM in the clean limit, we now consider the inclusion of the disorder, the form of which is given as,
\begin{equation}
H_{dis}=\sum_{r,s}\epsilon_{r,s} c^\dagger_{r,s} c_{r,s}
\end{equation}
where $r$ and $s$ are the coordinate in the lattice and spinor index respectively. $\epsilon_{r,s}$ is a uniformly distributed random number in the range of $[-W/2,W/2]$ utilized to mimic the random on-site disorder potential to be added into Eq. (\ref{eq:wham}). Note that this disorder configuration does not preserve time-reversal symmetry, therefore we are considering both charge and magnetic disorder. Within a given distribution of the disorder, we begin by specifying the correlation function of the disorder, whose average value zero, in the following form $\langle\epsilon_{i,s}\rangle=0$. The two point correlation of the disorder energy is given as, $\langle\epsilon_{i_1,s_1}\epsilon_{i_2,s_2}\rangle=\frac{[\int^{W/2}_{-W/2} \epsilon^2 d\epsilon]}{W}\delta_{i_1,i_2}\delta_{s_1,s_2}=\frac{W^2}{12}\delta_{i_1,i_2}\delta_{s_1,s_2}$. To more clearly see the effect of the disorder, we calculate the disorder averaged self-energy term in Green's function, $\bar{G}$, that is given as,
\begin{equation}
\bar{G}=\frac{1}{E-H_W-\Sigma_{dis}+i\eta}=\langle\frac{1}{E-(H_W+H_{dis})+i\eta}\rangle,
\end{equation}
where $<>$ indicates the average expectation value over the random disorder configurations and $\eta$ is the infinitesimal broadening term. To calculate the self energy, $\Sigma_{dis}$, we use the Dyson equation to dress the single particle Green's function given as,
\bea
\bar{G}=G_0+G_0\Sigma \bar{G}
\eea
where $G_0=\frac{1}{E-H_W+i\eta}$ is the Green's function of the bare Hamiltonian without disorder. Then, applying the Born approximation, the correction of the self-energy term is equivalent to\cite{bensimons},
\begin{gather}
\label{borngeneral}
\Sigma=\frac{[\int^{W/2}_{-W/2} \epsilon^2 d\epsilon]}{W} [I_2][<r_i |G_0 | r_i> ]
\\
\nonumber
=\frac{W^2}{12}\int d^3 k \frac{1}{E-H+i\eta}.
\end{gather}
\begin{figure}
\includegraphics[width=0.4\textwidth]{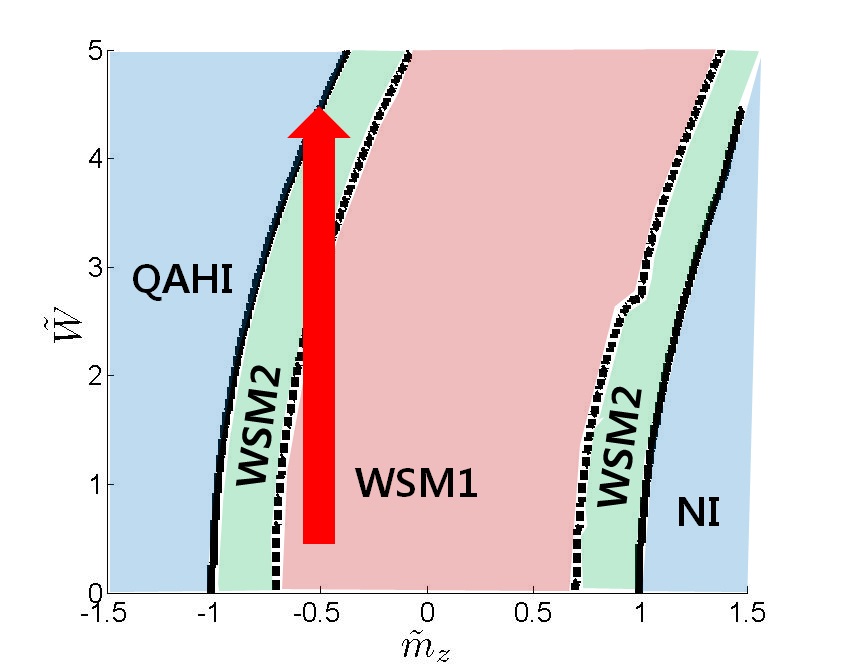}
\caption{\label{fig:sup1} 
The phase diagram of the Weyl semimetal with the effect of disorder included as calculated from the first Born approximation with the parameters, $\tilde{t}_x=\tilde{t}_y=\tilde{a}_{t}=1$. In this figure, we represent the different phases of WSM1, WSM2 and the insulating phase. We observe that, depending on the initial negative values of $\tilde{m}_z$, as we increase the disorder there exists a phase transition from WSM1 to WSM2 and finally to an insulating phase. However, we also note that for a range of values of positive $\tilde{m}_z$, we find that insulating metals may be driven through both the WSM2 and WSM1 phases with increasing disorder.}
\end{figure}
When we decompose the self-energy into its various directional contributions, these correspond to terms in which the chemical potential and the topological mass near $E=0$ are renormalized as\cite{PhysRevLett.115.246603,PhysRevLett.103.196805,bensimons},
\bea
\label{eq:renorm}
\mu_{re}=\mu+lim_{k\rightarrow0}Re(\Sigma_0),
\\
m_{i,re}=m_{i}+lim_{k\rightarrow0}Re(\Sigma_i).
\eea
Where $\Sigma_i$ is the self-energy correction is decomposed into each of the $i$-th directional pauli matrices. We observe that $Re(\Sigma_{x,y})$ vanishes since the numerator of the Eq. (\ref{borngeneral}) is an odd function of the momentum. Therefore, within the Born approximation, the value of the topological mass terms, $m_x$ and $m_y$, are invariant even with the disorder included in our analysis. Using Eq. (\ref{borngeneral}), $\Sigma_z$ is derived as,
\begin{gather}
\Sigma_z\approx+W^2\alpha.
\end{gather}
where the integral expression in the above equation, $\alpha$, is derived from the evaluation of the Eq. (\ref{borngeneral}) as,
\begin{widetext}
\begin{equation}
\label{Iexpression}
\alpha=
\\
-\frac{1}{12}\int \frac{d^3 k}{(2\pi)^3} \frac{m_0(2-cos(k_x) -cos(k_y)) +(m_z-t_z cos(k_z))}{|m_0(2-cos(k_x) -cos(k_y)) +(m_z-t_z cos(k_z))|^2+|t_x sin(k_x)|^2+|t_y sin(k_y)|^2-|a_{t} sin(k_z)+b_{t}cos(k_z)|^2}
\end{equation}
\begin{equation}
\label{Iapprox}
\approx\frac{-1}{48 m_0\pi}log(|\frac{m_0^2\pi^4}{(m_z+\sqrt{{m_z^2-(a_{t}^2+(b_{t}-t_z)^2)}})(m_z+\sqrt{{m_z^2-(a_{t}^2+(b_{t}+t_z)^2)}})}|).
\end{equation}
\end{widetext}
To obtain Eq. (\ref{Iapprox}), we have kept terms only up to quadratic order of $k_{x,y}$ contributions of the integral. In similar manner as outlined in Eq. (\ref{eq:renorm}), a renormalization of the chemical potential also occurs when $b_{tilt,i}\neq 0$ since the $b_{tilt}$ term is even function in the momentum space. However, the change of the chemical potential does not alter the phase of the WSM. We only focus on the renormalization of the topological mass.

In Eq. (\ref{Iapprox}), we notice that the sign of $\alpha$ is always negative (of $m_0$), therefore, the disorder renormalizes the value of $m_z$ to decreasing values as the magnitude of the disorder is increased as,
\bea
\label{massrenorm}
m_z \rightarrow m_z +\alpha W^2
\eea
Furthermore, the Fermi velocity in $z$ direction is also renormalized as
\bea
\label{velrenorm}
v_f=\tilde{t}_{z}sin(acos(m_z/t_z))
\\
\nonumber
 \rightarrow \tilde{t}_{z}sin(acos((m_z+\alpha W^2)/t_z))
\eea
which results in the renormalization of the effective Fermi velocity in $z$ direction. Therefore, while the tilt is invariant in the presence of the disorder, there is a decrease (increase) of the Fermi velocity when $\tilde{m}_z>0 (\tilde{m}_z<0)$. We find that when the effective Fermi velocity becomes smaller than the tilt at $\tilde{m}_z<0$, WSM1 undergoes the disorder induced quantum phase transition to WSM2 and then to an insulating phase with increasing disorder, as shown in Fig. \ref{fig:sup1} indicated by the arrow. Additionally, the reverse transition from an insulating phase to WSM2 and finally to WSM1 with increasing disorder is possible if we start from $\tilde{m}_z\approx 1$.

\textit{Numerical Calculation}-While the first-order Born approximation offers an analytical insight of the disorder induced phase transition, to confirm the result from the Born approximation, we numerically investigate the disorder effect via exact diagonalization of the three-dimensional tight-binding Hamiltonian of Eq. (\ref{eq:wham}).
\begin{figure}
\includegraphics[width=0.5\textwidth]{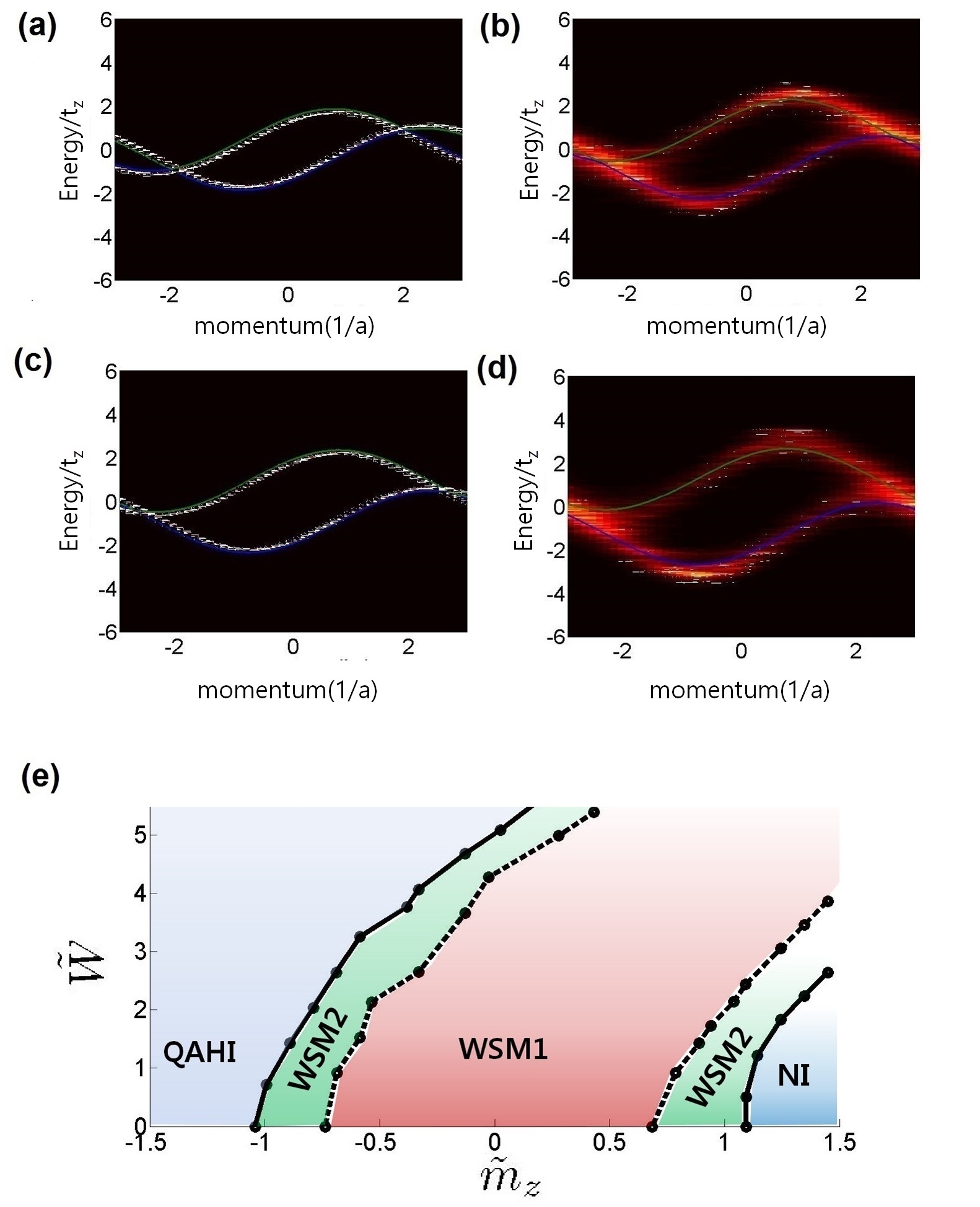}
\caption{\label{fig:numerics} Spectral function and disorder induced phase diagram derived from exact diagonalization. In each of the plots at the top (a),(b),(c) and (d), we show the numerically calculated spectral function as a function of $k_z$. The plots (a),(b),(c), and (d) use the following parameters for $(\tilde{m_z},\tilde{W}):(-0.27,0),(-0.27,3),(-0.78,0),(-0.78,4)$ respectively. We find that the system undergoes a phase transitions to WSM1-WSM2-QAHI, as predicted from the Born approximation. (e) The full phase diagram calculated from the exact diagonalization. This figure is generated with the parameters, $\tilde{t}_x=\tilde{t}_y=\tilde{a}_{t}=1$. The parameters used in (a)-(d) are marked with different shapes on (e) respectively.}
\end{figure}
In the presence of the disorder, we calculate the momentum space spectral function to observe the phase transitions with increasing disorder. The spectral function calculation gives an estimate of the tilt and the topological mass with a finite broadening of the states due to the presence of the disorder\cite{spectral}. As the value of the disorder increases, we use the largest value of the spectral function at a given $k_z$ to identify the resultant change of the dispersion. The spectral function in the momentum space is defined as\cite{datta},
\bea
A(\omega,\vec{k})=\sum_{i=1}^2 \frac{|\phi_{n,i}(\vec{k})|^2}{\omega-E_n+i\eta}
\eea
Where $\eta$ is the infinitesimal imaginary number and $i$ is the band index of the eigenstate. $E_n$ and $\phi_{n,i}(\vec{k})$ are the energy eigenvalue and the eigenstate of the Hamiltonian with disorder respectively. The spectral function is calculated using a system size of $8\times 8\times 60$ sites in the real space lattice. We numerically distinguish the WSM1, WSM2, and insulating phases by identifying the resultant changes of the dispersions and the tilt at the cones, while the change of the tilt is calculated from the slopes of the dispersions at the cone. We identify the insulating phases when the spectral function contains a gapped dispersion, and, similarly, we identify the WSM2 phase when the resultant dispersion shows the characteristic tilted cones. Fig.  \ref{fig:numerics} shows the disorder induced transition from the WSM1 to the WSM2 and, finally, to the insulating phase. To understand these different disorder induced transitions, we begin with Fig. \ref{fig:numerics} (a), which uses $(\tilde{m}_z,\tilde{W})=(-0.27,0)$, where we present the spectral function which contains degenerate crossings of two bands at $\vec{k}\approx(0,0,\pm 2)$ where the Weyl nodes are located in momentum space along the $\hat{z}$-direction. The spectral function has no broadening as the corresponding phase and dispersion are calculated in the clean limit. We find that the two Weyl cones have no tilt, which indicates WSM1 phase. Gradual increase of $W$ reduces the renormalized $m_z$ according to Eq. (\ref{massrenorm}), until $\gamma_{tilt}$ dominates the Fermi velocity and resulting in WSM2 phase. Eventually, when $\tilde{W}$ reaches up to 3 in Fig. \ref{fig:numerics} (b), the calculated dispersion shows the tilted Weyl cones along $\hat{z}$ direction, therefore it signals the disorder induced phase transition from WSM1 to WSM2. In addition to the WSM1-WSM2 phase transition, Fig. \ref{fig:numerics} (c) and (d) show a transition from WSM2 phase to insulating phase. Fig. \ref{fig:numerics} (c), with the choice of the parameters $(\tilde{m}_z,\tilde{W})=(-0.78,0)$ shows WSM2 phase in the clean limit in which the spectral function has no broadening and shows the tilted Weyl cones. Again, the increase of $W$ reduces the renormalized $m_z$ until when the location of the nodes, $Q=\pm(acos(\tilde{m}_z)$, are ill-defined to gap out the dispersion. Fig. \ref{fig:numerics} (d) shows this transition, as the disorder, $\tilde{W}$, reaches the value of 4. Fig. \ref{fig:numerics} (d) shows the spectral function of the insulating phase which are fully gapped out. In result, the behavior of the transitions shown in the numerical calculation can be understood from the decrease of $m_z$ and the resulting change of the Fermi velocity due to the disorder induced renormalization, in which it eventually shifts the full phase diagram to the positive $m_z$ direction.

In Fig. \ref{fig:numerics} (e), we show the complete disorder induced phase diagram of the WSM as a function of $\tilde{W}$ and $\tilde{m}_z$ in which we derive by repeating the calculations with various values of $m_z$. The numerical phase diagram agrees in its behavior with the phase transitions predicted using the Born approximation in Fig. \ref{fig:sup1}. The general trend, in which $m_z$ reduces as the disorder increases, is similar to the previous study\cite{PhysRevLett.115.246603}. However, in contrast to the previous study, we find that there always exists a finite region of WSM2 phase before the WSM1 phase undergoes the transition to an insulating phase, when $a_{tilt}$ is non-zero. This intermediate region of the WSM2 phase is guaranteed to exist because the Fermi velocity eventually vanishes and the tilt dominates before the insulating transition. Due to the WSM1-WSM2 transition, WSM1 phase near $\tilde{m}_z=-cos(atan(a_{t}))$ is unstable to the arbitrary weak disorder to the transition to the WSM2 phase. On the opposite side of the phase diagram where $m_z$ is positive, WSM2 phase near $\tilde{m}_z=cos(atan(a_{t}))$ is unstable to the WSM1 phase transition as the renormalized $m_z$ decreases the Fermi velocity. On both sides of the phase boundaries, the WSM2 phase is unstable to the finite disorder, resulting the transitions to WSM1($\tilde{m}_z= cos(atan(a_{t}))$) and QAHI ($\tilde{m}_z= -1$), as the renormalized Fermi velocity and $m_z$ is significantly modified. As the disorder increases large enough, Anderson localization can occur where the perturbation theory fails and the bulk gap collapses\cite{PhysRevLett.115.246603}.

\textit{Conclusion}-
In conclusion, we have studied the effect of disorder on the resultant phase diagram of WSM with a non-zero tilt to elucidate the boundaries of the different physical regimes as a function of disorder and topological mass. We have illustrated these various phase transitions both analytically, using the first Born approximation, and numerically, via exact diagonalization calculations. We find that the renormalization of the topological mass changes the effective Fermi velocity of the nodes reducing both the mass and resultant Fermi velocity with increasing values of disorder. The resulting change of the Fermi velocity leads to phase transitions between WSM1, WSM2, and insulator. Moreover, our results show that the disorder induced WSM2 phase always occurs before the metal-insulator transition of WSM1. Therefore, we assert that the WSM2 phase naturally occurs before the disorder induced transition between WSM1 and the insulating phases in the known WSM materials.
\section{Acknowledgement}
This work is supported by NSF CAREER ECCS-1351871.
\bibliography{disorderbib}
\pagebreak
\onecolumngrid

\end{document}